
\input jytex.tex   
\typesize=10pt
\magnification=1200
\baselineskip=17truept
\hsize=6truein\vsize=8.5truein
\sectionnumstyle{blank}
\chapternumstyle{blank}
\chapternum=1
\sectionnum=1
\pagenum=0

\def\begintitle{\pagenumstyle{blank}\parindent=0pt\begin{narrow}[0.4in]}
\def\endtitle{\end{narrow}\newpage\pagenumstyle{arabic}}


\def\beginexercise{\vskip 20truept\parindent=0pt\begin{narrow}[10
truept]}
\def\endexercise{\vskip 10truept\end{narrow}}


\def\eql#1{\eqno\eqnlabel{#1}}
\def\ref{\reference}
\def\peq{\puteqn}
\def\pref{\putref}

\def\mgn{\marginnote}
\def\bex{\begin{exercise}}
\def\eex{\end{exercise}}

\def\mbox#1{{\leavevmode\hbox{#1}}}
\def\hspace#1{{\phantom{\mbox#1}}}

\def\al{\alpha}
\def\be{\beta}
\def\ga{\gamma}

\def\Ga{\Gamma}

\def\ep{\epsilon}

\def\ze{\zeta}


\def\frac#1/#2{\leavevmode\kern.1em
\raise.5ex\hbox{\the\scriptfont0 #1}\kern-.1em/\kern-.15em
\lower.25ex\hbox{\the\scriptfont0 #2}}
\def\sfrac#1/#2{\leavevmode\kern.1em
\raise.5ex\hbox{\the\scriptscriptfont0 #1}\kern-.1em/\kern-.15em
\lower.25ex\hbox{\the\scriptscriptfont0 #2}}

\def\gtorder{\mathrel{\raise.3ex\hbox{$>$}\mkern-14mu
             \lower0.6ex\hbox{$\sim$}}}
\def\ltorder{\mathrel{\raise.3ex\hbox{$<$}\mkern-14mu
             \lower0.6ex\hbox{$\sim$}}}

\def\semidirprod{\rlap{\ss C}\raise1pt\hbox{$\mkern.75mu\times$}}
\def\for{\lower6pt\hbox{$\Big|$}}
\def\fish{\kern-.25em{\phantom{abcde}\over \phantom{abcde}}\kern-.25em}


\def\boxit#1{\vbox{\hrule\hbox{\vrule\kern3pt
        \vbox{\kern3pt#1\kern3pt}\kern3pt\vrule}\hrule}}
\def\dalemb#1#2{{\vbox{\hrule height .#2pt
        \hbox{\vrule width.#2pt height#1pt \kern#1pt
                \vrule width.#2pt}
        \hrule height.#2pt}}}


\def\noin{\noindent}


\def\eg{{\it e.g. }}
\def\ie{{\it i.e. }}


  %

\def\3j#1#2#3#4#5#6{\left\lgroup\matrix{#1&#2&#3\cr#4&#5&#6\cr}
\right\rgroup}

\def\m?{\mgn{?}}


\def\aop#1#2#3{{\it Ann. Phys.} {\bf {#1}} (19{#2}) #3}

\def\cmp#1#2#3{{\it Comm. Math. Phys.} {\bf {#1}} (19{#2}) #3}
\def\cqg#1#2#3{{\it Class. Quant. Grav.} {\bf {#1}} (19{#2}) #3}

\def\jmp#1#2#3{{\it J. Math. Phys.} {\bf {#1}} (19{#2}) #3}
\def\jpa#1#2#3{{\it J. Phys.} {\bf A{#1}} (19{#2}) #3}

\def\np#1#2#3{{\it Nucl. Phys.} {\bf B{#1}} (19{#2}) #3}
\def\pl#1#2#3{{\it Phys. Lett.} {\bf {#1}} (19{#2}) #3}
\def\pm#1#2#3{{\it Phil.Mag.} {\bf {#1}} ({#2}) #3}

\def\pr#1#2#3{{\it Phys. Rev.} {\bf {#1}} (19{#2}) #3}
\def\prA#1#2#3{{\it Phys. Rev.} {\bf A{#1}} (19{#2}) #3}

\def\prD#1#2#3{{\it Phys. Rev.} {\bf D{#1}} (19{#2}) #3}

\def\prs#1#2#3{{\it Proc. Roy. Soc.} {\bf A{#1}} (19{#2}) #3}
\def\pcps#1#2#3{{\it Proc. Camb. Phil. Soc.} {\bf{#1}} (19{#2}) #3}

\def\dmj#1#2#3{{\it Duke Math. J.} {\bf {#1}} (19{#2}) #3}

\def\jdg#1#2#3{{\it J. Diff. Geom.} {\bf {#1}} (19{#2}) #3}
\def\jfa#1#2#3{{\it J. Func. Anal.} {\bf {#1}} (19{#2}) #3}

\def\ma#1#2#3{{\it Math. Ann.} {\bf {#1}} ({#2}) #3}
\def\mz#1#2#3{{\it Math. Zeit.} {\bf {#1}} ({#2}) #3}
\def\pams#1#2#3{{\it Proc. Am. Math. Soc.} {\bf {#1}} (19{#2}) #3}

\def\qjm#1#2#3{{\it Quart. J. Math.} {\bf {#1}} (19{#2}) #3}

\def\tams#1#2#3{{\it Trans. Am. Math. Soc.} {\bf {#1}} (19{#2}) #3}

\begin{title}
\vglue 20truept
\righttext {MUTP/95/13}
\righttext{hep-th/95}
\leftline{\today}
\vskip 100truept
\centertext {\Bigfonts \bf Spin on the 4-ball}
\vskip 15truept
\centertext{J.S.Dowker\footnote{Dowker@a3.ph.man.ac.uk}}
\vskip 7truept
\centertext{\it Department of Theoretical Physics,\\
The University of Manchester, Manchester, England.}
\vskip 60truept
\centertext {Abstract}
\begin{narrow}
Using known mode properties, the functional determinant for massless
spin-half fields on the Euclidean 4-ball is calculated and shown to be
different for spectral (nonlocal) and mixed (local) boundary conditions.
The local result agrees with that from a conformal argument. Some
higher-spin results and a sum rule are also given.
\end{narrow}
\vskip 5truept
\righttext {August 1995}
\vskip 75truept
\righttext{Typeset in \jyTeX}
\vfil
\end{title}
\pagenum=0
\section{\bf 1. Introduction}
The theory of spinors in spaces with boundaries is of interest physically in
connection with quantum cosmology and supergravity. (See D'Eath and Esposito
[\pref{DandE}] and Esposito [\pref{Esposito}] for some history of these
questions.) In mathematics it is encountered in the spin-index theorem and the
Atiyah, Patodi and Singer $\eta$ spectral asymmetry function, the standard
reference being Gilkey's book, [\pref{gilk}].

As explained in [\pref{DandE}], for self-adjointness of the Dirac operator,
there is a choice between spectral and local (mixed) boundary conditions,
the former being of relevance for the spin-index and the latter having more
physical significance in connection with supersymmetry, string theory
and quantum gravity, [\pref{MandP2,Luck}], although in the guise of relative
conditions they do have a cohomological importance, [\pref{gilk,BandG2}].

In the special case of the Euclidean 4-ball, it was shown [\pref{DandE2,Kam,
KandM}] that the value of $\ze(0)$, which determines the scaling of the theory,
was the same for both sets of conditions. In this note we report on the
same question for the one-loop effective action, which is, up to factors,
$\ze'(0)$. Our method will be that explained in [\pref{Dow9}].
\section{\bf 2. Mode properties and calculation}
The analysis of the modes of the massless Dirac equation on the 4-ball
was carried out by D'Eath and Esposito [\pref{DandE,DandE2}] and we will do
no more here on this matter than use their results. For local
boundary conditions they found that the eigenvalues, $\al^2$, are the roots of
the equation
$$ F^L_p(\al)=J_{p-1}^2(\al)-J_p^2(\al)=0
\eql{locb}$$with a degeneracy, for a given $p$, of $p^2-p$, $p=1,2,\ldots$.
For spectral conditions, there is the simpler, scalar-like condition,
$$
F_p^S(\al)=J_p(\al)=0
\eql{specb}$$
with degeneracy $2(p^2+p)$, $p=1,2,\ldots$.

Our approach is based on the Mittag-Leffler decomposition,
$$
z^{-\be}F_p(z)=\ga\prod_\al\bigg(1-{z^2\over\al^2}\bigg),
\eql{ml1}$$
where
$$\eqalign{
&\be=p,\quad\ga={1\over2^pp!},\quad{\rm spectral},\cr
&\be=2(p-1),\quad\ga={1\over\big(2^{p-1}(p-1)!\big)^2},\quad{\rm local}.
\cr}
$$

This standard decomposition was earlier employed by Moss [\pref{Moss}] and by
D'Eath and Esposito [\pref{DandE}] when looking at the heat-kernel expansion
and $\ze(0)$. Here, when finding $\ze'(0)$, we need the normalising factor,
$\ga$, which follows from the small-$z$ behaviour of $F_p(z)$.

A few details of the calculation will be given but, for brevity,
some of our previous work must be utilised.

Bypassing a number of steps, which are fully explained in [\pref{Dow9,Dow8}],
we define the quantities
$$
G_N\sim\sum_{p=1}^\infty p^N\bigg[\big(p-{1\over2}\big)
\ln{2p\over p+\ep}+(\ep-p)+\sum_{n=1}^{N+1}\bigg({E_n(t)\over \ep^n}
-{E_n(1)\over p^n}\bigg)+I_N(p)\bigg]
\eql{logdetl}$$
and
$$
H_N\sim\sum_{p=1}^\infty p^N\bigg[p\ln{2p
\over p+\ep}+\ep-p-{1\over2}\ln{\ep\over p}+\sum_{n=1}^{N+1}
\bigg({T_n(t)\over \ep^n}-{T_n(1)\over p^n}\bigg)+I_N(p)\bigg],
\eql{logdets}$$
with
$$
I_N(p)=\int_0^\infty\bigg({1\over2}
-{1\over\tau}+\sum_{k=1}^{[N/2]+1}(-1)^kB_{2k}{\tau^{2k-1}\over(2k)!}
+{1\over e^\tau-1}\bigg){e^{-\tau p}\over\tau}\,d\tau,
$$
in terms of which we can write the spin-half quantities,
$$\eqalign{
&{\ze_{1/2}^L}'(0)=2\big(G_2-G_1\big),\cr
&{\ze^S_{1/2}}'(0)=2\big(H_2+H_1\big).\cr}
\eql{zedashes}$$
The labels $S$ and $L$ refer to spectral and local boundary conditions
respectively.

In equations (\peq{logdetl}) and (\peq{logdets}) the $\sim$ symbol signifies
that the mass-independent part of the large-mass asymptotic limit is to be
taken. The $E_n(t)$ are the polynomials in $t=p/\ep$, $\ep=(m^2+p^2)^{1/2}$,
that occur in the asymptotic expansion of $F_p^L(im)$ of (\peq{locb}) derived
by D'Eath and Esposito (they call them $A_n/2$) from Olver's series. The
$T_n(t)$ are the corresponding polynomials for the scalar case,
[\pref{Moss,Dow8}]. The condition that makes equation (\peq{logdetl})
possible is $E_n(1)=T_n(1)$ which can be proved from the explicit definition
of the $E_n$. We note that $T_n(1)$ is zero for $n$ even and that
$T_{2k-1}(1)=(-1)^kB_{2k}/2k(2k-1)$ in terms of Bernoulli numbers.

We have made use of the algebraic results of D'Eath and Esposito,
[\pref{DandE}] section IV, in deriving (\peq{logdetl}).

Expression (\peq{logdets}) is identical to one occurring for scalar fields
on the even ball, except that $N$, there being the power of
$p$ in the expansion of the degeneracy, is even. Hence for $N=2$,
our previous result in [\pref{Dow8,DandA2}] for the 4-ball (see also
[\pref{BGKE}]) could be used without change.

{}From the technique outlined in [\pref{Dow9}] the following useful limits
can be deduced,\mgn{work out ln m part}
$$\eqalign{
&\sum_{p=1}^\infty p^N(\ep-p)\sim -\ze_R(-N-1)+O(\ln m),\cr
&\sum_{p=1}^\infty p^N\ln\big({2p\over p+\ep}\big)\sim-\ze_R'(-N)+\ln2\,
\ze_R(-N)+O(\ln m),\cr
&\sum_{p=1}^\infty p^N\ln\big({\ep\over p}\big)\sim\ze_R'(-N)+O(\ln m).\cr}
\eql{aslims}$$

It is necessary to state that a hidden regularisation has been employed to
render the summations finite. This consists of removing sufficient of the
Taylor expansion of the summand and will not be indicated. Since the entire
expression is finite, the divergent terms so introduced must all cancel.

These limits enable some of the terms in (\peq{logdetl}) and (\peq{logdets})
to be dealt with quickly. The rest, \ie the polynomial and integral
contributions, need a little more work. We write them as in
[\pref{Dow8,Dow9}],
$$\eqalign{
\sum_{p=1}^\infty&\, p^N\bigg[\sum_{n=1,3,\ldots}^{N+1}
P_n(1) \bigg({1\over\ep^n}-{1\over p^n}\bigg)
+\sum_{n=1}^{N+1}{P'_n(t)\over\ep^n}\bigg]\cr
&+\lim_{s\to0}\!\int_0^\infty\!\!\bigg({1\over2}\!
-\!{1\over\tau}\!+\!\sum_{k=1}^{[N/2]+1}\!\!(-1)^kB_{2k}{\tau^{2k-1}
\over(2k)!}\!
+\!{1\over e^\tau-1}\bigg)\tau^{s-1}(-1)^N{d^N\over d\tau^N}
{1\over e^\tau\!-\!1}d\tau,\cr}
\eql{rem2}$$ where $P_n$ stands for either $E_n$ or $T_n$ and $P'_n(t)=P_n(t)
-P_n(1)$.

A recursion is developed for the multiple derivative in (\peq{rem2}) and the
contribution from the integral found to be, after some algebra,
$$
\ze_R'(-N-1)+{1\over2}\ze_R'(-N)+\ze_R(-N-1)+
\sum_{k=1}^{N+1}M_k^{(N)}\ze_R'(-k),
\eql{intcont}$$
where the coefficient matrix $M$ is defined by
$$
M_k^{(N)}=\sum_{l=k}^{N+1}A_l^{(N)}{S_{l+1}^{(k+1)}\over l!}
$$
in terms of easily evaluated recursion constants $A_l^{(j)}$ and Stirling
numbers $S_l^{(k)}$, [\pref{Dow8,Dow9}].

Assembling the various pieces, and using special values for the $M_k^{(N)}$,
we find
$$\eqalign{
G_N&={\ze_R'(-N)\over2}\!+\!{\ze_R'(-N-1)\over N+1}\!
+\!\sum_{k=1}^{N-1}\!M_k^{(N)}\ze_R'(-k)\!\cr
&+\!\big({1\over2}\ze_R(-N)-\ze_R(\!-\!N\!-1)\big)\,\ln2
+\int_0^1t^NE''_{N+1}(t)\,dt+L_N,\quad N\ge1,\cr}
\eql{totl}$$
where $P_n''(t)=P'_n(t)/(1-t^2)$ and
$$
L_N=T_{N+1}(1)\bigg(\ln2
+\sum_{k=1}^N{1\over k}+\sum_{q=1}^{N/2}
{(-1)^q\sqrt\pi\,(N/2)!\over2q(N/2-q)!\Ga(q+1/2)}\bigg).
$$
The last two terms in (\peq{totl}) come from the first line of (\peq{rem2}).

Explicitly for $N=0$, a case needed later,
$$
G_0=-{1\over24}+{1\over12}\ln2+\ze_R'(-1).
\eql{gee0}$$

For spectral conditions,
$$\eqalign{
H_N&=-{\ze_R'(-N)\over2}+{\ze_R'(-N-1)\over N+1}
+\sum_{k=1}^{N-1}M_k^{(N)}\ze_R'(-k)\cr
&+\ze_R(-N-1)\,\ln2
+\int_0^1t^NT''_{N+1}(t)\,dt+L_N,\quad N\ge1,\cr}
\eql{tots}$$
and
$$
H_0= {5\over24}-{1\over6}\ln2+\ze_R'(-1)-\ze_R'(0).
\eql{aitch0}$$

Making the constructions (\peq{zedashes}), one finds for local spin-half,
\mgn{SPINHALF.MTH}
$$
{\ze^L_{1/2}}'(0)={251\over15120}-{11\over180}\ln2
+{2\over3}\big(\ze_R'(-3)-\ze_R'(-1)\big)\approx0.088108
\eql{dashl}$$
and for spectral,
$$
{\ze^S_{1/2}}'(0)=-{2489\over30240}+{1\over45}\ln2
+{2\over3}\big(\ze_R'(-3)-\ze_R'(-1)\big)\approx0.046962
\eql{dashs}$$
which are the main results of this note.

The specific forms of the $E_n$ polynomials given in [\pref{DandE}], have
been used to
evaluate the integrals in (\peq{totl}). We remark that in the corresponding
evaluation of $\ze(0)$ (=$11/360$), one needs only the particular value
$P_N(1)$, which equals $\ze_R(-N)/N$, a non-transcendental, local quantity.

\section{\bf 3. Higher spins}
The eigenvalue conditions for some higher-spin theories are summarised in
[\pref{DandE2}] section VI. A mechanical application of the present technique
yields the following results.

For real spin-0 with Dirichlet conditions,
$$\eqalign{
2{\ze^D_0}'(0)&=2H_2\cr
\noalign{\vskip5truept}
&={173\over15120}+{1\over45}\ln2+{2\over3}\ze_R'(-3)-\ze_R'(-2)
+{1\over3}\ze_R'(-1)\cr
\noalign{\vskip5truept}
&\approx0.005738.\cr}
$$
For spin-1 (Maxwell) with Dirichlet (magnetic) conditions,
$$\eqalign{
\ze'_{\rm TV}(0)&=2(H_2-2H_0)\cr
\noalign{\vskip5truept}
&=-{6127\over15120}+{16\over45}\ln2+{2\over3}
\ze_R'(-3)-\ze_R'(-2)-{5\over3}\ze_R'(-1)+2\ze_R'(0)\cr
\noalign{\vskip5truept}
&\approx-1.68691.\cr}
\eql{TV}$$
For spin-3/2 physical degrees of freedom with spectral conditions,
$$\eqalign{
{\ze^S_{3/2}}'(0)&=2(H_2+H_1-H_0)\cr
\noalign{\vskip5truept}
&=-{27689\over30240}+{31\over45}\ln2+{2\over3}\ze_R'(-3)
-{14\over3}\ze_R'(-1)+4\ze_R'(0)\cr
\noalign{\vskip5truept}
&\approx-3.33834.\cr}
\eql{3/2s}$$
These results imply, rather trivially, the sum rule,
$$
{\ze^S_{3/2}}'(0)-{\ze^S_{1/2}}'(0)=2\big(\ze_{\rm TV}'(0)
-2{\ze^D_0}'(0)\big).
\eql{ident1}$$

The same relation holds also for $\ze(0)$,
$$
{\ze^S_{3/2}}(0)-{\ze^S_{1/2}}(0)=2\big(\ze_{\rm TV}(0)
-2\ze^D_0(0)\big),
\eql{ident2}$$
and, in fact, for all coefficients in the heat-kernel expansion, as can be
checked numerically from the tables provided in [\pref{BEK}] and
[\pref{KandC}].

The specific values,
$$
\ze^S_{3/2}(0)=-{289\over360},\quad\ze^S_{1/2}(0)={11\over360},
\quad\ze_{\rm TV}(0)=-{77\over180},\quad \ze^D_0(0)=-{1\over180},
\eql{zezeros}
$$
were computed in references [\pref{DandE,DandE2,Louko}], see also
[\pref{MandP,KandC,Poletti}]. The spectral label, $S$, can be replaced
by the local one, $L$, in (\peq{zezeros}).

The sum rules are only special cases of the general relation
$$
{\ze^S_{3/2}}(s)-{\ze^S_{1/2}}(s)=2\big(\ze_{\rm TV}(s)
-2\ze^D_0(s)\big),
\eql{ident3}$$
which is a consequence of the eigenvalue condition, (\peq{specb}), and
the various quadratic degeneracies.

For spin-2 transverse-traceless modes with Dirichlet conditions,
[\pref{Schleich}], \ \ie
$$
F_p^{\rm TT}=J_p(\al)=0
$$ and degeneracy $2(p^2-4)$, $p\ge3$, we find
$$\eqalign{
\ze'_{\rm TT}(0)&=2(\overline H_2-4\overline H_0)=2(H_2-H_0)+
6\big(\ze_R'(0)+\ln2\big)\cr
&= -{25027\over15120}+{331\over45}\ln2+{2\over3}\ze_R'(-3)-\ze_R'(-2)
-{23\over3}\ze_R'(-1)+14\ze_R'(0)\cr
\noalign{\vskip5truept}
&\approx-8.119619,\cr}
\eql{TT}$$
where the bar signifies that the $p=1$ term has been left out in
(\peq{logdets}). (The easiest way of doing this is to remove the overall $p=1$
term at the outset.)

For the record, the local spin-3/2 expression is
$$\eqalign{
{\ze^L_{3/2}}'(0)&=2(\overline G_2-\overline G_1-2\overline G_0)\cr
&={2771\over15120}+{289\over180}\ln2+{2\over3}\ze_R'(-3)
-{14\over3}\ze_R'(-1)+4\ze_R'(0)\cr
\noalign{\vskip5truept}
&\approx-1.60405,\cr}
\eql{3/2l}$$
which exhibits the anomaly value of $-289/360$.

Arbitrary-spin fields can be treated in exactly the same way, most easily
using the mode analysis given in [\pref{Dow10,DandC2}], and will be discussed
in a later communication.

\begin{ignore}

For massless spin-j fields, the degeneracy is $2(n^2-j^2)$, [\pref{Dow10}],
where $n=p$ for integer spins with Dirichlet conditions and $n=p+j$ for
half odd-integer ones with spectral conditions.
\end{ignore}

\section{\bf 4. Comments}
The above expressions for the $\ze'(0)$ have also been obtained by Kirsten
and Cognola [\pref{KandC}] using the method of Bordag {\it et al},
[\pref{BGKE}].

The local result, (\peq{dashl}), agrees with that of Apps, reported
in [\pref{DandA2}] and found using a
conformal transformation from the 4-hemisphere. In fact, the final expression
in (\peq{dashl}) is $\ze'_S(0)$ on the hemisphere, the rest coming from the
cocycle function obtained from an integration of the conformal anomaly, as in
[\pref{DandA,BandG}] for example.

Spectral conditions are also conformally invariant and it seems that
(\peq{dashs}) can be interpreted in a similar way. The same structure is also
apparent in (\peq{3/2s}) and (\peq{3/2l}) for spin 3/2.

This suggests that the eigenvalue problem on the hemisphere is the same,
or is equivalent, for spectral and local boundary conditions. This is
confirmed
by, and may explain, the equality of $\ze(0)$ for these conditions found
by D'Eath and Esposito in flat space and by Kamenshchik and
Mishakov on the bounded sphere. To the author's knowledge, the cocycle
function has not been calculated for spectral conditions.

The extension to higher, even-dimensional spaces is straightforward and simply
consists of substituting (\peq{totl}) or (\peq{tots}) into the appropriate
polynomial form of the spinor degeneracy. For odd dimensions the major
difference is that the $p$-sums run over half odd-integers and presents no
problem [\pref{Dow9}]. For example, the Maxwell modes on the 3-ball are
classic, \eg [\pref{deB}], and it is soon shown that the magnetic determinant
is obtained by doubling the scalar Dirichlet value and subtracting $-2\ln2$
to allow for the different starting point of the mode sum. Similarly, the
electric determinant is the double the scalar Robin one, with $\beta=1/2$,
again minus $-2\ln2$.
\section{\bf Acknowlegment}
I wish to thank Klaus Kirsten for helpfully communicating his results.

\vskip 10truept
\noin{\bf{References}}
\vskip 5truept
\begin{putreferences}
\ref{Rayleigh}{Lord Rayleigh{\it Theory of Sound} vols.I and II,
MacMillan, London, 1877,78.}
\ref{KCD}{G.Kennedy, R.Critchley and J.S.Dowker \aop{125}{80}{346}.}
\ref{Donnelly} {H.Donnelly \ma{224}{1976}161.}
\ref{Fur2}{D.V.Fursaev {\sl Spectral geometry and one-loop divergences on
manifolds with conical singularities}, JINR preprint DSF-13/94,
hep-th/9405143.}
\ref{HandE}{S.W.Hawking and G.F.R.Ellis {\sl The large scale structure of
space-time} Cambridge University Press, 1973.}
\ref{DandK}{J.S.Dowker and G.Kennedy \jpa{11}{78}{895}.}
\ref{ChandD}{Peter Chang and J.S.Dowker \np{395}{93}{407}.}
\ref{FandM}{D.V.Fursaev and G.Miele \pr{D49}{94}{987}.}
\ref{Dowkerccs}{J.S.Dowker \cqg{4}{87}{L157}.}
\ref{BandH}{J.Br\"uning and E.Heintze \dmj{51}{84}{959}.}
\ref{Cheeger}{J.Cheeger \jdg{18}{83}{575}.}
\ref{SandW}{K.Stewartson and R.T.Waechter \pcps{69}{71}{353}.}
\ref{CandJ}{H.S.Carslaw and J.C.Jaeger {\it The conduction of heat
in solids} Oxford, The Clarendon Press, 1959.}
\ref{BandH}{H.P.Baltes and E.M.Hilf {\it Spectra of finite systems}.}
\ref{Epstein}{P.Epstein \ma{56}{1903}{615}.}
\ref{Kennedy1}{G.Kennedy \pr{D23}{81}{2884}.}
\ref{Kennedy2}{G.Kennedy PhD thesis, Manchester (1978).}
\ref{Kennedy3}{G.Kennedy \jpa{11}{78}{L173}.}
\ref{Luscher}{M.L\"uscher, K.Symanzik and P.Weiss \np {173}{80}{365}.}
\ref{Polyakov}{A.M.Polyakov \pl {103}{81}{207}.}
\ref{Bukhb}{L.Bukhbinder, V.P.Gusynin and P.I.Fomin {\it Sov. J. Nucl.
 Phys.} {\bf 44} (1986) 534.}
\ref{Alvarez}{O.Alvarez \np {216}{83}{125}.}
\ref{DandS}{J.S.Dowker and J.P.Schofield \jmp{31}{90}{808}.}
\ref{Dow1}{J.S.Dowker \cmp{162}{94}{633}.}
\ref{Dow2}{J.S.Dowker \cqg{11}{94}{557}.}
\ref{Dow3}{J.S.Dowker \jmp{35}{94}{4989}; erratum {\it ibid}, Feb.1995.}
\ref{Dow5}{J.S.Dowker {\it Heat-kernels and polytopes} To be published}
\ref{Dow6}{J.S.Dowker \pr{D50}{94}{6369}.}
\ref{Dow7}{J.S.Dowker \pr{D39}{89}{1235}.}
\ref{Dow8}{J.S.Dowker {\it Robin conditions on the Euclidean ball}
MUTP/95/7; hep-th\break/9506042.}
\ref{Dow9}{J.S.Dowker {\it Oddball determinants} MUTP/95/12; hep-th/9507096.}
\ref{Dow10}{J.S.Dowker \pr{D28}{83}{3013}.}
\ref{BandG}{P.B.Gilkey and T.P.Branson \tams{344}{94}{479}.}
\ref{Schofield}{J.P.Schofield Ph.D.thesis, University of Manchester,
(1991).}
\ref{Barnesa}{E.W.Barnes {\it Trans. Camb. Phil. Soc.} {\bf 19} (1903)
374.}
\ref{BandG2}{T.P.Branson and P.B.Gilkey {\it Comm. Partial Diff. Equations}
{\bf 15} (1990) 245.}
\ref{Pathria}{R.K.Pathria {\it Suppl.Nuovo Cim.} {\bf 4} (1966) 276.}
\ref{Baltes}{H.P.Baltes \prA{6}{72}{2252}.}
\ref{Spivak}{M.Spivak {\it Differential Geometry} vols III, IV, Publish
or Perish, Boston, 1975.}
\ref{Eisenhart}{L.P.Eisenhart {\it Differential Geometry}, Princeton
University Press, Princeton, 1926.}
\ref{Moss}{I.G.Moss \cqg{6}{89}{659}.}
\ref{Barv}{A.O.Barvinsky, Yu.A.Kamenshchik and I.P.Karmazin \aop {219}
{92}{201}.}
\ref{Kam}{Yu.A.Kamenshchik and I.V.Mishakov \prD{47}{93}{1380}.}
\ref{KandM}{Yu.A.Kamenshchik and I.V.Mishakov {\it Int. J. Mod. Phys.}
{\bf A7} (1992) 3265.}
\ref{DandE}{P.D.D'Eath and G.V.M.Esposito \prD{43}{91}{3234}.}
\ref{DandE2}{P.D.D'Eath and G.V.M.Esposito \prD{44}{91}{1713}.}
\ref{Rich}{K.Richardson \jfa{122}{94}{52}.}
\ref{Osgood}{B.Osgood, R.Phillips and P.Sarnak \jfa{80}{88}{148}.}
\ref{BCY}{T.P.Branson, S.-Y. A.Chang and P.C.Yang \cmp{149}{92}{241}.}
\ref{Vass}{D.V.Vassilevich.{\it Vector fields on a disk with mixed
boundary conditions} gr-qc /9404052.}
\ref{MandP}{I.Moss and S.Poletti \pl{B333}{94}{326}.}
\ref{Kam2}{G.Esposito, A.Y.Kamenshchik, I.V.Mishakov and G.Pollifrone
\prD{50}{94}{6329}.}
\ref{Aurell1}{E.Aurell and P.Salomonson \cmp{165}{94}{233}.}
\ref{Aurell2}{E.Aurell and P.Salomonson {\it Further results on functional
determinants of laplacians on simplicial complexes} hep-th/9405140.}
\ref{BandO}{T.P.Branson and B.\O rsted \pams{113}{91}{669}.}
\ref{Elizalde1}{E.Elizalde, \jmp{35}{94}{3308}.}
\ref{BandK}{M.Bordag and K.Kirsten {\it Heat-kernel coefficients of
the Laplace operator on the 3-dimensional ball} hep-th/9501064.}
\ref{Waechter}{R.T.Waechter \pcps{72}{72}{439}.}
\ref{GRV}{S.Guraswamy, S.G.Rajeev and P.Vitale {\it O(N) sigma-model as
a three dimensional conformal field theory}, Rochester preprint UR-1357.}
\ref{CandC}{A.Capelli and A.Costa \np {314}{89}{707}.}
\ref{IandZ}{C.Itzykson and J.-B.Zuber \np{275}{86}{580}.}
\ref{BandH}{M.V.Berry and C.J.Howls \prs {447}{94}{527}.}
\ref{DandW}{A.Dettki and A.Wipf \np{377}{92}{252}.}
\ref{Weisbergerb} {W.I.Weisberger \cmp{112}{87}{633}.}
\ref{Voros}{A.Voros \cmp{110}{87}{110}.}
\ref{Pockels}{F.Pockels {\it \"Uber die partielle Differentialgleichung
$\Delta u+k^2u=0$}, B.G.Teubner, Leipzig 1891.}
\ref{Kober}{H.Kober \mz{39}{1935}{609}.}
\ref{Watson2}{G.N.Watson \qjm{2}{31}{300}.}
\ref{DandC1}{J.S.Dowker and R.Critchley \prD {13}{76}{3224}.}
\ref{DandC2}{J.S.Dowker and R.Critchley \prD {13}{76}{224}.}
\ref{Lamb}{H.Lamb \pm{15}{1884}{270}.}
\ref{EandR}{E.Elizalde and A.Romeo International J. of Math. and Phys.
{\bf13} (1994) 453}
\ref{DandA}{J.S.Dowker and J.S.Apps \cqg{12}{95}{1363}.}
\ref{DandA2}{J.S.Dowker and J.S.Apps, {\it Functional determinants on certain
domains}. To appear in the Proceedings of the 6th Moscow Quantum Gravity
Seminar held in Moscow, June 1995; hep-th/9506204.}
\ref{Watson1}{G.N.Watson {\it Theory of Bessel Functions} Cambridge
University Press, Cambridge, 1944.}
\ref{BGKE}{M.Bordag, B.Geyer, K.Kirsten and E.Elizalde, {\it Zeta function
determinant of the Laplace operator on the D-dimensional ball} UB-ECM-PF
95/10; hep-th /9505157.}
\ref{MandO}{W.Magnus and F.Oberhettinger {\it Formeln und S\"atze}
Springer-Verlag, Berlin, 1948.}
\ref{Olver}{F.W.J.Olver {\it Phil.Trans.Roy.Soc} {\bf A247} (1954) 328.}
\ref{Hurt}{N.E.Hurt {\it Geometric Quantization in action} Reidel,
Dordrecht, 1983.}
\ref{Esposito}{G.Esposito {\it Quantum Gravity, Quantum Cosmology and
Lorentzian Geometry}, Lecture Notes in Physics, Monographs, Vol. m12,
Springer-Verlag, Berlin 1994.}
\ref{Louko}{J.Louko \prD{38}{88}{478}.}
\ref{Schleich} {K.Schleich \prD{32}{85}{1989}.}
\ref{BEK}{M.Bordag, E.Elizalde and K.Kirsten {\it Heat kernel
coefficients of the Laplace operator on the D-dimensional ball}
 UB-ECM-PF 95/3; hep-th/9503023.}
\ref{ELZ}{E.Elizalde, S.Leseduarte and S.Zerbini.}
\ref{BGV}{T.P.Branson, P.B.Gilkey and D.V.Vassilevich {\it The Asymptotics
of the Laplacian on a manifold with boundary} II, hep-th/9504029.}
\ref{Erdelyi}{A.Erdelyi,W.Magnus,F.Oberhettinger and F.G.Tricomi {\it
Higher Transcendental Functions} Vol.I McGraw-Hill, New York, 1953.}
\ref{Quine}{J.R.Quine, S.H.Heydari and R.Y.Song \tams{338}{93}{213}.}
\ref{Dikii}{L.A.Dikii {\it Usp. Mat. Nauk.} {\bf13} (1958) 111.}
\ref{DandH}{P.D.D'Eath and J.J.Halliwell \prd{35}{87}{1100}.}
\ref{KandC}{K.Kirsten and G.Cognola, {\it Heat-kernel coefficients and
functional determinants for higher spin fields on the ball} UTF354. Aug. 1995.}
\ref{Louko}{J.Louko \prD{38}{88}{478}.}
\ref{MandP}{I.G.Moss and S.J.Poletti \pl{B333}{94}{326}.}
\ref{MandP2}{I.G.Moss and S.J.Poletti \np{341}{90}{155}.}
\ref{Luck}{H.C.Luckock \jmp{32}{91}{1755}.}
\ref{Poletti}{S.J.Poletti \pl{B249}{90}{355}.}
\ref{gilk}{P.B.Gilkey {\it Invariant theory, the heat equation and the
Atiyah-Singer index theorem}, Publish or Perish, Wilmington, DE, 1984.}
\ref{deB}{Louis de Broglie {\it Probl\`emes de propagation guide\'es des
ondes electromagnetiques} 2me. \'Ed. Gauthier-Villars, Paris, 1951.}
\end{putreferences}
\bye